\documentclass[prl,showpacs,byrevtex]{revtex4}
\input epsf.tex
\def\DESepsf(#1 width #2){\epsfxsize=#2 \epsfbox{#1}}
\usepackage{graphicx}
\usepackage{epsfig}
\usepackage{rotating}
\usepackage{amsmath}
\usepackage{amsfonts}
\usepackage{amssymb}
\usepackage{url}
\usepackage{hyperref}
\usepackage[footnotesize]{caption2}
\usepackage{bm}
\def\ba{\begin{eqnarray}}
\def\ea{\end{eqnarray}}
\def\br{\begin{array}}
\def\er{\end{array}}
\def\be{\begin{equation}}
\def\ee{\end{equation}}
\def\uoo{c_{13}c_{12}}
\def\uot{c_{13}s_{12}}
\def\uoth{s_{13}e^{-i\delta}}
\def\uto{-c_{23}s_{12}-c_{12}s_{13}s_{23}e^{i\delta}}
\def\utt{c_{12}c_{23}-s_{12}s_{13}s_{23} e^{i\delta}}
\def\utth{c_{13}s_{23}}
\def\utho{s_{12}s_{23}-c_{12}s_{13}c_{23}e^{i\delta}}
\def\utht{-c_{12}s_{23}-c_{23}s_{13}s_{12} e^{i\delta}}
\def\uthth{c_{13}c_{23}}

\newcommand\bi{\begin{itemize}}
\newcommand\ei{\end{itemize}}

\def\sc{\sin\theta_{13}}
\def\sa{\sin\theta_{23}}
\def\cs{\cos\theta_{12}}

\def\ca{\cos\theta_{23}}

\def\da1{d\alpha_1\over dt}
\def\da2{d\alpha_2\over dt}
\def\ovl{\overline}
\def\cj{{\cal J}}
\def\ck{{\cal K}}

\def\cn{{\cal N}}
\begin{document}
\thispagestyle{empty}
\begin{center}
{ \large\bf {Inverse Seesaw Mechanism in Nonsupersymmetric
$SO(10)$, Proton Lifetime, Nonunitarity Effects, and a Low-mass
 $Z^{\prime}$ Boson}}
\vskip .15in
{\bf Ram Lal Awasthi ${}^{\dag}$ and Mina K. Parida ${}^{*}$}\\

{\sl${}^*${{\em Centre of Excellence in Theoretical and Mathematical
~Sciences,\\
{\rm SOA} University,
 Khandagiri Square, Bhubaneswar  751030, India}}}\\ 

{\sl${}^{\dag}${{\em Harish-Chandra Research Institute, Chhatnag Road, Jhusi,\\
    Allahabad 211019, India}}\\} 
\end{center}

\begin{abstract}
Recently realization of TeV scale inverse seesaw mechanism   
 in supersymmetric $SO(10)$ framework has led to a number of experimentally
verifiable
predictions including low-mass $W_R^{\pm}$ and $Z^{\prime}$
 gauge bosons and nonunitarity effects. Using 
nonsupersymmetric $SO(10)$ grand unified theory, we show how
a TeV scale inverse seesaw mechanism for neutrino masses is implemented with 
 a low-mass
 $Z^{\prime}$ boson accessible to Large Hadron Collider. We derive
 renormalization group equations for fermion masses and mixings in the
 presence of the  intermediate symmetries of the model and extract the Dirac
 neutrino mass matrix at the TeV scale from successful GUT-scale parameterization of
 fermion masses. We estimate leptonic
 nonunitarity effects measurable at neutrino factories and lepton flavor
 violating decays expected to be probed in near future. While our prediction 
on the nonunitarity matrix element $\eta_{\mu\tau}$ for degenerate right-handed neutrinos 
is similar to the supersymmetric $SO(10)$ case, we find new predictions
with significantly enhanced value of 
 its phase $\delta_{\mu\tau}\simeq 10^{-4}-10^{-2}$ when partial degeneracy 
among these neutrino masses is adequately taken into account by a constraint
 relation that emerges naturally in this approach. Other  
 predictions on branching ratios and CP-violating parameters are discussed.
  An important distinguishing characteristic as another test of the 
best identified minimal model
 is that the threshold corrected two-loop prediction on proton lifetime with maximum value $(\tau_p)_{\rm
 max.}\simeq 10^{35}$ yrs. is accessible to ongoing search experiments for
 the  decay $p\to e^+\pi^0$.  
 Simple model extensions 
 with longer proton lifetime predictions are also discussed.
\end{abstract}
\maketitle

\section{I.INTRODUCTION}

 Supersymmetric grand unified theories (GUTs) provide
 a very attractive framework for representing particles and forces of nature as they
 solve the gauge hierarchy problem, unify three forces of nature, and also
 explain tiny 
neutrino masses through seesaw paradigm \cite{nurev}  while providing possible cold dark
 matter candidates of the universe.
 An evidence of
 supersymmetry at the Large Hadron Collider (LHC) would be a land-mark 
 discovery  which would certainly change the future course of
 physics. 
But, in the absence of any evidence of supersymmetry so far, it is worth while to explore
new physics prospects of
nonsupersymmetric (non-SUSY) GUTs \cite{ps,su5,so10} and, particularly, those based upon $SO(10)$ which has grown in
popularity as it unifies all fermions of one generation including the
 right-handed (RH) neutrino into a single spinorial representation. It provides
 spontaneous origins of P ($=$ Parity) and CP violations \cite{lrs,dpar1,dpar2}. Most interestingly, in
 addition to predicting the right
 order of tiny neutrino masses through mechanisms 
called the canonical ($\equiv$ type-I) \cite{type-I} seesaw and type-II
\cite{type-II} seesaw, it has high
 potentiality to explain all fermion masses \cite{baburnm,rnm1} including large 
mixings in the neutrino sector \cite{bsv} with type-II seesaw dominance \cite{t2dom,melfo,alta}.  
In fact neither seesaw mechanism, nor grand unification require supersymmetry
 per se.
Although gauge couplings automatically unify in the minimal supersymmetric
 standard model \cite{goran}, and they fail to unify through the minimal particle content of
 the standard model (SM) in one-step breaking of non-SUSY SU(5) or SO(10), they do unify once 
intermediate symmetries are included to populate the grand desert in case of
 non-SUSY SO(10) \cite{dpar2,pal,mp,lmpr}. In addition,
 with intermediate gauge
 symmetries $SO(10)$ also predicts signals of new physics which can be probed
 at low or accelerator energies.\\

A hallmark of $SO(10)$ grand unification is its underlying quark-lepton
symmetry \cite{ps} because of which the canonical seesaw scale is pushed closer to the 
GUT scale making it naturally inaccessible to direct tests by low-energy experiments or collider searches. The energy scale of type-II seesaw mechanism in $SO(10)$ is also too high for direct experimental tests. In contrast to these high scale seesaw mechanisms, an experimentally verifiable and attractive  mechanism that has been 
recently introduced into  $SO(10)$ \cite{mkprad} is the  radiative seesaw \cite{ma} where the quark-lepton unification has no role to play and 
additional suppression to light neutrino mass prediction occurs by loop mediation proportional to a small Higgs quartic coupling that naturally emerges from a Plank-scale induced term in the GUT Lagrangian. 
The model predicts a rich structure of prospective dark matter candidates also verifiable by ongoing search experiments. It has been further noted that this embedding of the radiative seesaw in $SO(10)$ may have a  promising prospect for representing all fermion masses. 
A number of other interesting neutrino mass generation mechanisms including type-III seesaw, double seesaw, linear seesaw, scalar- triplet seesaw have been suggested and some of them are also experimentally verifiable \cite{nurev}.

In the context of non-SUSY $SO(10)$ in this work our purpose is to explore the prospects of another neutrino mass generation mechanism called the inverse seesaw \cite{rnmv} which 
is also different from canonical or Type-II seesaw mechanism and has the potentiality 
to be experimentally verifiable because of the low scale at which it can operate  although 
higher scale inverse seesaw  models  have been  
suggested \cite{smirnov,pr}. In a large class of models \cite{naturalness,n2, khalil,gs}, the implementation requires the introduction of fermionic singlets under the gauge group of the model.  
Likewise, its implementation in $SO(10)$ introduces a new mass scale $\mu_S$ into the
Lagrangian corresponding to the mass matrix of the additional singlet fermions of three generations and the TeV-scale seesaw  requires this parameter to be small. There is an interesting naturalness argument in favor of its smallness 
based upon exact lepton number conservation symmetry \cite{naturalness,psb1}. Below the TeV scale, in the limit ${\mu}_S \to 0$, the corresponding Lagrangian has a leptonic global $U(1)$ symmetry which guarantees left-handed neutrinos to remain massless. The small value of ${\mu}_S$  essentially needed to match the neutrino oscillation data 
with the TeV-scale inverse seesaw formula may be taken as a consequence of very mild breaking of the exact global symmetry. Thus the small value of the parameter protected by the exact lepton number conservation is natural in the 't Hooft sense \cite{gthooft}. However, inspite of such  
interesting naturalness argument,
there has been no dynamical understanding of its origin so far although an interpretation using Higgs mechanism has been given in the context of a model with extended gauge, fermion and Higgs sectors \cite{ma2}. In a different class of  $SO(10)$ models,       
the small singlet fermion mass parameter has been generated radiatively \cite{invrad} where more nonstandard fermions have been found to be necessary. In SUSY $SO(10)$ singlet fermions have also been used to derive new forms of fermion  mass matrix while predicting standard fermion mass ratios \cite{barr1} and to obtain new seesaw formula for neutrinos while explaining baryon asymmetry of the universe through leptogenesis \cite{barr2}.
While most of the inverse seesaw models need gauge singlet fermions under the
 SM gauge group or its extensions \cite{rnmv,smirnov,pr,naturalness,n2,khalil,gs,psb1,ma2,invrad} and the use of $SO(10)$-singlet fermions may point to the disadvantages of the corresponding GUT-based models, extended electroweak theory based upon $SU(3)_L\times SU(3)_R\times U(1)$ gauge symmetry \cite{dias331} contains such singlets in its fundamental representations. To give some examples of GUTs, one $SO(10)$-singlet fermion per generation is automatically contained in the $27-$ dimensional fermion representation of  $E_{6}$
\cite {gursey} where $27=16+10+1$ under $SO(10)$ but $27=(3,3^*,1)+(3,1,3^*)+(1,3,3^*)$ under $SU(3)_L\times
SU(3)_R\times SU(3)_C$  and, in the latter case, an additional discrete $Z_3$ symmetry is needed to qualify it as a trinification GUT model \cite{trini}.  Interesting properties of $SU(3)^3$ gauge theory including experimentally verifiable predictions  at accelerator energies have been 
discussed \cite{cpwsu33,masu33}. Gauge boson mediated proton decays are suppressed in $SU(3)^3$ type of models. In addition to the RH neutrino and the other singlet fermion needed for inverse seesaw, these models ( $SU(3)^3$ and $E_6$)
also contain $10$ nonstandard fermions per generation and no experimental data are yet available on their masses at low energies so as to persue the present bottom-up approach  to derive the Dirac neutrino mass matrix from fermion mass fits at the GUT scale. The same argument holds against any other model that may contain additional
 nonstandard fermions beyond the RH neutrino and the singlet-fermion needed for inverse seesaw.

Regarding the potentiality of SO(10) motivated inverse seesaw in the visible sector, the same quark-lepton symmetry that forces
canonical seesaw scale to be far beyond the experimentally accessible range, makes the
TeV-scale inverse seesaw predict observable nonunitarity effects as new physics
signals verifiable at low and accelerator energies and at neutrino factories \cite{psb1}.

Recently in a series of interesting investigations, using inverse seesaw mechanism,
Bhupal Dev and Mohapatra \cite{psb1}      
have shown that SUSY $SO(10)$, besides admitting a low spontaneous breaking scale of
$SU(2)_L\times SU(2)_R\times U(1)_{B-L} \times SU(3)_C(\equiv G_{2213})$ gauge symmetry 
 with right handed gauge bosons 
$W_R^{\pm}$ and $Z^{\prime}$ accessible to LHC, is also capable of fitting all fermion masses
and mixings at the GUT scale  while predicting 
observable nonunitarity effects. The model has been also shown to account for the observed baryon asymmetry of the universe through leptogenesis caused due to the decay of TeV scale masses of the pseudo Dirac RH neutrinos \cite{psb2}. Currently considerable attention has been
devoted to propose models with an extra neutral $Z^{\prime}$ gauge boson 
which may also emerge from Pati-Salam or left-right gauge theories, or
$SO(10)$ and $E_6$ grand unified theories, or also from string inspired
models \cite{zprev}. 

Different from the supersymmetric $SO(10)$
model of ref.\cite{psb1}, here we adopt the view that there may not be any manifestation of supersymmetry at accelerator energies and that the actual parity restoration scale may be
high. Instead of both the $W_R^{\pm}$ and the $Z^{\prime}$ boson masses being low,
there may be only some remnants of high scale left-right symmetry or quark
lepton symmetry manifesting at low and accelerator energies as smoking
gun signatures such as the $Z^{\prime}$ \cite{zprev,lhczp} gauge boson and the associated 
nonunitarity effects of the TeV-scale inverse seesaw. With this point of view 
in this work we show that a non-SUSY $SO(10)$ with $SU(2)_L\times U(1)_R\times 
U(1)_{(B-L)}\times SU(3)_C (\equiv G_{2113})$ gauge symmetry at the TeV scale and left-right gauge 
theory at higher intermediate scale, with or without D-parity, achieves precision 
gauge coupling unification, and predicts a low mass $Z^{\prime}$ making them suitable 
for implementation of TeV-scale inverse seesaw mechanism. The model can also be verified or falsified through 
its predictions on observable nonunitarity effects and additional contributions to lepton
flavor violations. Another testing ground for the model could be through the $SO(10)$ prediction on gauge boson mediated proton decay on which dedicated search experiments are ongoing.

 We derive renormalization group equations in the presence
of two intermediate gauge symmetries for running fermion masses and mixings,
and determine the Dirac neutrino mass at the TeV scale from successful fits to
the fermion masses at the GUT scale. In this approach we find a simple relation
between the RH neutrino masses in the model. We also point out a different type of relation in the partial degenerate case that permits much lower values of RH neutrino masses resulting in  a  CP-violating
phase increased by $2-4$ orders larger than the degenerate case. Some of our predictions 
include branching ratios for $\mu\to e\gamma$ enhanced by $1-2$ orders. Out of the two 
minimal models, while the intermediate scale D-Parity conserving model is ruled out by 
proton decay constraint, the proton lifetime for $p\to e^+\pi^0$ in the intermediate 
scale D-Parity nonconserving model is predicted to be well within the accessible range 
of ongoing search experiments. We have also discussed simple extensions of the two models
with longer proton lifetime predictions. This method can also be implemented using 
Pati-Salam model or left-right models \cite{ps,lrs}.

This paper is organized in the following manner. In Sec.2 we briefly
discuss the model and carry out gauge coupling unification and proton
lifetime predictions in Sec.3. With a brief explanation of inverse seesaw
mechanism in Sec.4 we summarize relevant formulas encoding nonunitarity
effects and lepton flavor violations. In Sec.5 we discuss renormalization group evolution 
of fermion masses and mixings to the GUT scale in the presence of nonsupersymmetric gauge 
theories $G_{2113}$ and $G_{2213}$ at intermediate scales. In this section we also show 
how fermion masses are fitted at the GUT scale and information on the Dirac neutrino mass
matrix is obtained. Nonunitarity effects are discussed in Sec.6 with  predictions on the
moduli of relevant matrix elements. In Sec.7 we give predictions on CP-violating parameters and lepton flavor violation where we also discuss possible limitations of the present models.
In Sec.8 we provide a brief summary and discussion along with conclusion. In the Appendix A we provide beta function coefficients for gauge coupling 
unification while in Appendix B we summarize derivations of renormalization group equations 
(RGEs) for fermion masses and mixings.

\section{II.THE MODEL}

There has been extensive investigation on physically appealing intermediate
scale models \cite{dpar1,dpar2,pal,lmpr,mal1} in non-SUSY $SO(10)$.
Although in the minimal two step-breaking of non-SUSY $SO(10)$ models \cite{lmpr} we found no
suitable chain with a sufficiently low scale to implement the inverse seesaw,
the following chain with two intermediate gauge symmetries appears to be quite suitable,
\ba
SO(10) &
\stackrel {(M_U)}
{\longrightarrow} 
& G_I \nonumber \\
&
\stackrel {(M_R^+)}
{\longrightarrow}& SU(2)_L \times
U(1)_R \times U(1)_{B-L}\times  SU(3)_C~~[G_{2113}] \nonumber \\
&
\stackrel {(M_R^0)}
{\longrightarrow}& SU(2)_L \times
U(1)_Y\times  SU(3)_C~~[\rm SM] \nonumber\\
&
\stackrel {(M_Z)}{\longrightarrow}&SU(3)_C \times U(1)_Q ,\label{chain}
\ea
where we will consider two possibilities for $G_I$.\\ 

As Model-I $G_I=SU(2)_L \times
SU(2)_R \times U(1)_{B-L}\times  SU(3)_C ~~[\equiv G_{2213}](g_{2L} \neq
g_{2R})$  is realized by breaking the GUT-symmetry and by assigning vacuum
expectation value (VEV) to the D-Parity odd singlet in ${45}_H$ \cite{dpar1}. 
As the left-right discrete symmetry is spontaneously broken at the GUT scale, 
the Higgs sector becomes asymmetric below $\mu=M_U$ causing inequality between 
the gauge couplings $g_{2L}$ and $g_{2R}$. This model does not have the cosmological domain wall problem. The second step of symmetry breaking 
takes place by the right-handed (RH) Higgs triplet $\sigma_R(1,3,0,1)\subset {45}_H$ 
whereas the third step of breaking to SM takes place by the $G_{2113}$-submultiplet 
$\chi_R^0(1,1/2, -1/2,1)$ contained in the RH doublet of ${16}_H$. It is well known 
that SM breaks to low energy symmetry by the SM Higgs doublet contained in the 
 bidoublet $(2,2,0,1)$ under $G_{2213}$ which originates from ${10}_H$ of $SO(10)$.        
This is the minimal particle content for the model to carry out the spontaneous 
braking of GUT symmetry to low-energy theory. But a major objective of the present 
work is to explore the possibility of observable nonunitarity effects for which it 
is required to extract information on the Dirac neutrino mass matrix ($M_D$) from a 
fit to the fermion masses at the GUT scale and this is possible by including two Higgs
doublets instead of one \cite{psb1}. We assume  these doublets to
originate from two separate bidoublets contained in ${10}_H^a(a= 1,2)$. 
Implementation of inverse seesaw also requires the minimal extension 
 by adding three $SO(10)$-singlet fermions $S_i (i=1, 2, 3)$, one for each generation \cite{rnmv}.

As Model-II, we treat the GUT  symmetry to be broken by the VEV of the $G_{2213}$-singlet 
$(1,1,0,1)\subset {210}_H$ which is even under D-parity \cite{dpar1}. This causes 
the Higgs sector below GUT-scale to be left-right symmetric resulting in 
equal gauge couplings in $G_I=G_{2213D}(g_{2L}= g_{2R})$. For the sake of simplicity we 
treat the rest of the symmetry breaking patterns of Model-II similar to Model-I and we 
assume the presence of three singlet fermions.
We call these two models, Model-I and Model-II, as minimal models with two low scale  Higgs 
doublets in each. We now examine precision gauge coupling unification for these two models.

\section{III.GAUGE COUPLING UNIFICATION AND PROTON LIFETIME}

In this section we examine gauge coupling unification in the minimal Model-I and
and the minimal Model-II and make predictions on proton lifetimes while we  also predict the
corresponding quantities in their simple extensions.

\begin{center}
{\bf A. Unification in minimal models}
\end{center}
 
\begin{figure}[htbp]
\centering
\includegraphics[width=0.45\textwidth,height=0.45\textheight,angle=-90]
{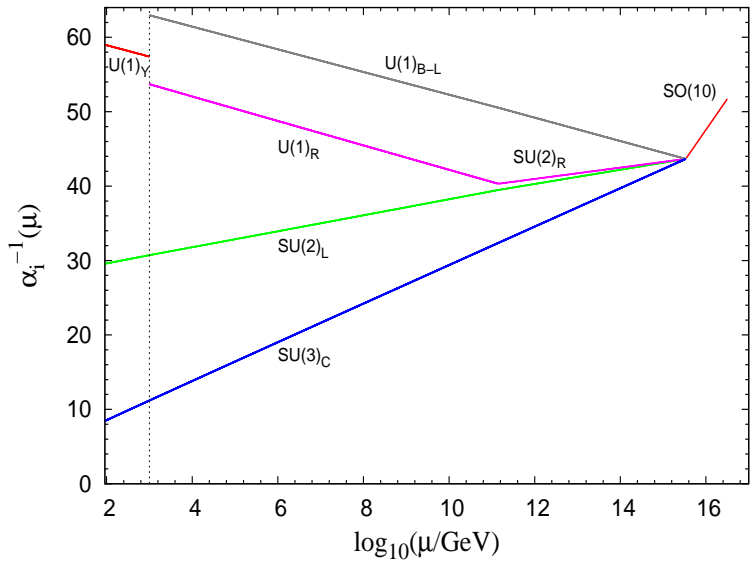}
\caption{Gauge coupling unification in Model-I with two-loop values
  $M_U=10^{15.53}$ GeV and $M_{R^+}= 10^{11.15}$ GeV with a low mass
  $Z^{\prime}$ gauge boson at $M_{R^0} \sim 1$ TeV.}   
\label{Fig1}
\end{figure}

It was shown in \cite{dpar2} that with  $G_{2113}$ gauge symmetry at the lowest intermediate scale
in $SO(10)$  there is substantial impact of two-loop effects on mass scale predictions  in a number 
of cases. The one-loop and the two-loop beta-function coefficients for the evolution of  gauge
couplings \cite{gqw,jones} for Model-I and Model-II with two Higgs doublets for
each case are given in  Appendix A. We have also included small mixing effects \cite{mal1,mix} due to two 
abelian gauge factors $U(1)_R\times U(1)_{(B-L)}$ in both the models below the $M_R^+$ scale. Using $\sin^2\theta_W(M_Z)=0.23116\pm 0.00013,
~\alpha^{-1}(M_Z)=127.9$ and $\alpha_S(M_Z)=0.1184\pm 0.0007$ \cite{nakamura} 
we find that with $M_{Z^{\prime}}\sim M_{R^0} \sim 1$ TeV precision unification of gauge couplings 
occurs for the following values of masses at one-loop and two-loop levels for 
the Model-I,
\ba
M_U^{ol}&=&10^{15.978}~{\rm GeV}, M^{ol}_{R^+}=10^{10.787} ~{\rm GeV},~\alpha_G^{ol}=0.02253,~({\rm
  one-loop}),\nonumber\\
M_U&=&10^{15.530}~{\rm GeV}, M_{R^+}=10^{11.15} ~{\rm GeV},~\alpha_G=0.02290~({\rm
  two-loop}).\label{solmod1}
\ea
The RG evolution of gauge couplings at two-loop level is shown in Fig.
\ref{Fig1} exhibiting precision unification at $M_U= 10^{15.53}$ GeV. In Model-II
coupling unification occurs with similar precision but at $M_U=10^{15.17}$ GeV.

The decay width of the proton for $p\to e^+\pi^0$ is \cite{nath}
 \ba   
\Gamma(p\rightarrow e^+\pi^0) 
&=&\frac{m_p}{64\pi f_{\pi}^2}
\left(\frac{{g_G}^4}{{M_U}^4}\right)|A_L|^2|\bar{\alpha_H}|^2(1+D+F)^2\times R.
\label{width}
\ea
where  $R=[(A_{SR}^2+A_{SL}^2) (1+ |{V_{ud}}|^2)^2]$ for $SO(10)$, $V_{ud}=0.974=$ the  
$(1,1)$ element of $V_{CKM}$ for quark mixings, $A_{SL}(A_{SR})$ is the short-distance 
renormalization factor in the left (right) sectors and $A_L=1.25=$ long distance 
renormalization factor. $M_U=$ degenerate mass of $24$ superheavy gauge bosons in
$SO(10)$, $\bar\alpha_H =$ hadronic matrix element, $m_p = $ proton mass $=938.3$ MeV, 
$f_{\pi}=$ pion decay constant $=139$ MeV, and the chiral Lagrangian parameters are 
$D=0.81$ and $F=0.47$. With $\alpha_H= \bar{\alpha_H}(1+D+F)=0.012$ GeV$^3$ estimated 
from lattice gauge theory computations, we obtain  $A_R \simeq A_LA_{SL}\simeq
A_LA_{SR}\simeq 2.726$ for Model-I. The expression for the inverse decay rates for both 
the minimal models is expressed as, 
\ba
\Gamma^{-1}(p\rightarrow e^+\pi^0)
&=& (1.01\times 10^{34} {\rm Yrs})\left({0.012 GeV^3}\over
\alpha_H\right)^2
\left({2.726 \over A_R}\right)^2\left({{1/43.6}\over
\alpha_G}\right)^2\nonumber \\
&&\times\left({7.6\over F_q}\right) \left({M_U \over {2.98\times 10^{15} {\rm
GeV}}}\right)^4, \label{lifetime}
\ea
where the factor  $F_q=2(1+|V_{ud}|^2)^2\simeq 7.6$ for $SO(10)$. Now using  the estimated 
values of the model parameters in each case the predictions on proton lifetimes for 
both models are given in Table \ref{tab1} where the uncertainties in unification scale and proton
lifetime have been estimated by enhancing the error in $\alpha_S$ to $3\sigma$
level. It is clear
  that with maximal value $(\tau_p)_{max.}=7\times 10^{34}$ Yrs., Model-I predicts 
the proton lifetime closer to the current experimental lower bound  
${(\tau_p)}_{expt.}(p \to e^+\pi^0) \ge 1.2\times
  10^{34}$~~Yrs. \cite{nishino} which is accessible to ongoing proton decay
  searches in near future \cite{hyperk}. On the other hand Model-II is ruled out at two-loop
  level as it predicts lifetime nearly two orders smaller. The reduction of
  lifetime by nearly two-orders compared to one-loop predictions in both cases is
  due to the corresponding reduction in the unification scale by a factor
 of $\simeq 1/3$.   
 
\begin{table*}
\begin{center}
\begin{tabular}{|c|c|c|c|c|c|c|c|c|}
\hline
$Model$&$M^{ol}_U$&$M^{ol}_{R^+}$&$M_U$&$M_{R^+}$&$\alpha_G^{-1}$&${\rm A}_R$&$\tau^o_p$&$\tau_p$\\
      &(GeV)&(GeV)&(GeV)&(GeV)&&&$(yrs.)$&$(yrs.)$\\ \hline
$I$&$10^{15.978}$&$10^{10.787}$&$10^{15.530}$&$10^{11.150}$&$43.67$&$2.726$&$1.08\times
10^{36\pm 0.32}$&$2\times 10^{34\pm 0.32}$ \\ \hline
$II$&$10^{15.56\pm 0.08}$&$10^{11.475}$&$10^{15.17\pm 0.08}$&$10^{11.750}$&$42.738$&$2.670$&$
2.44\times 10^{34\pm 0.32}$&$6.3\times 10^{32\pm 0.32}$ \\ \hline
\end{tabular}
\end{center}
\caption{GUT scale, intermediate scale and proton lifetime predictions for
nonsupersymmetric SO(10) models with TeV scale $Z^{\prime}$ boson and two
Higgs doublets as described in the text. The uncertainty in the proton
lifetime has been estimated using $3\sigma$ uncertainty in $\alpha_S(M_Z)$.}
\label{tab1}
\end{table*}

The fact that the Model-I admits a low $(B-L)$ breaking scale corresponding to 
a light $Z^{\prime}$ accessible to accelerator searches makes this non-SUSY
model suitable to accommodate inverse seesaw mechanism. Unlike the SUSY $SO(10)$ 
model \cite{psb1}, here the $W_R^{\pm}$ bosons are far beyond  the LHC accessible range.

\begin{center}
{\bf B. Unification in simple model extensions}
\end{center}

Although the minimal Model-I clearly satisfies the proton decay constraint to accommodate 
TeV scale seesaw, we study simple extensions of both models to show that they can
evade proton lifetime constraint in case future  experiments show $\tau_p$ to 
be substantially longer than $10^{35}$ Yrs. We use an additional real color octet scalar 
$C_8(1,0,8)\subset {45}_H$ where the  quantum numbers are under the SM gauge group and allow its mass to vary between $1$ TeV and the GUT scale. 
Making it light would require additional fine tuning of parameters. Recently such a light 
scalar has been used in models with interesting phenomenological consequences and if the 
particle mass is in the accessible range, it may be produced at LHC with new physics signatures beyond the standard model \cite{dobrescu}.   
 
\begin{figure}[htbp]
\centering
\includegraphics[width=0.45\textwidth,height=0.45\textheight,angle=-90]{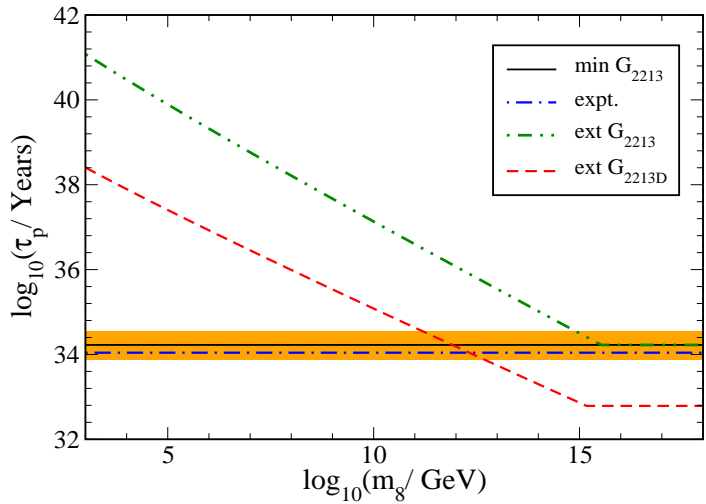}
\caption{Variation of proton lifetime as a function of  color octet mass in
  simple extensions of Model-I~(double dot-dashed line) and Model-II (dashed
  line). The horizontal solid line with error
  band is the prediction of the minimal
Model-I while the horizontal dot-dashed line is
  the experimental lower bound for $p\to e^+\pi^0$.} 
\label{Fig2}
\end{figure}

The presence of this scalar octet with lower mass makes the evolution of $\alpha_{3c}^{-1}(\mu)$ 
flatter thereby pushing the GUT scale to higher values. In Fig \ref{Fig2} we plot predicted proton 
lifetimes in the extended $G_{2213}$ and $G_{2213D}$ models as a function of the octet mass $m_8$. 
It is clear that such a simple extension of the two models can easily satisfy  proton lifetime requirements in             
future experimental measurements even if they are
found to be much longer than the current limit.

Then while the minimal Model-I can be easily chosen for inverse seesaw, both the models 
with such simple extension and possessing  TeV scale $U(1)_{(B-L)}$ breaking scale qualify 
for the same purpose.

\section{IV.INVERSE SEESAW AND  FORMULAS FOR CP AND LEPTON FLAVOR VIOLATIONS}

For the phenomenological study of nonunitarity effects we confine to the 
Model-I and all our analyses are similar for Model-II. Introducing additional
$SO(10)$-singlet fermions (S) for three generations, 
the Yukawa Lagrangian at the GUT scale gives rise to the
effective Lagrangian near the second intermediate scale $\mu = {\rm M}_R^0\sim
1$ TeV, 
\ba
{\cal L}_{\rm Yuk}&=&Y^a{\bf 16}\cdot{\bf 16}\cdot{\bf {10}_H}^a+y_{\chi} {\bf 16}\cdot{\bf
  1}\cdot{\bf {16}_H}^{\dagger}+{\bf \mu}_S {\bf 1}\cdot{\bf 1}\nonumber\\ 
&\supset& \left ( Y^a \ovl \psi_L \psi_R \Phi^a  + y_{\chi}\ovl \psi_RS\chi_R^0 
+ H.c.\right )+ S^T{ \mu}_S S, \label{yuklm}
\ea
where  the first (second) equation is invariant under $SO(10)$ ($G_{2113}$)  gauge  symmetry. The left-handed (LH) and the right-handed (RH) fermion fields 
$\psi_{\rm L} (2,0, -1/2,1)$, 
 $\psi_{\rm R} (1,1/2, -1/2,1)$ 
with their respective quantum numbers under $G_{2113}$ are contained in the spinorial representation ${\bf {16}}\subset SO(10)$ and the two Higgs doublets 
$\Phi (2,\pm 1/2,0,1)\subset {10}_H \subset SO(10)$.
 The Lagrangian has a new mass scale ${\mu}_S$ corresponding 
to the mass matrix of the $SO(10)$-singlet fermions. 
Denoting the RH neutrino mass as $ M_R = y_{\chi}v_{{\chi}}$ where
$v_{\chi}=\langle \chi^0_R\rangle $
and the Dirac mass matrix for neutrino as $M_D= Y_{\nu}v_u$ where $v_u$ is the VEV of the up-type Higgs doublet, eq.(\ref{yuklm}) gives the mass part of the neutrino sector in the Lagrangian in the flavor basis 
after the symmetry breaking $G_{2113}\to SM$ 
\ba
{\cal L}_{\rm mass}&=&\left({\bar {\nu}}M_{\rm D}N+{\bar N}M_RS+H.c.\right )+
 S^T{ \mu}_S S,\label{masslag}
\ea
which, in the $(\nu, N, S)_L$ basis, leads to a mass 
matrix \cite{rnmv,psb1}
\ba
{\cal M}_{\nu} =  
\begin{pmatrix} 0 & M_D & 0  \\
M_D^T & 0  & M_R\\ 0 & M_R^T & {\bf {\mu}_S}  \end{pmatrix}.  \label{matrix}
\ea
 Denoting
${\rm X}=M_DM_R^{-1}$, block diagonalization of eq.(\ref{matrix})
under the condition $M_R\gg M_D\gg {\mu}_S$ 
leads to the inverse seesaw formula for
light neutrino mass matrix,
\ba
{ m}_{\nu} &=&~ M_DM_R^{-1}{\bf \mu}_S (M_R^T)^{-1}M_D^T \equiv 
{\rm X}{\bf \mu}_S {\rm X}^T. 
\label{inv}
\ea
It is clear that the TeV-scale inverse seesaw formula is tenable and appropriate to fit the light neutrino masses provided ${\mu}_S$ 
is the smallest of the three mass scales occurring in eq.(\ref{matrix}).
 Based upon symmetry, there exist interesting naturalness arguments 
 in the literature in favor of smallness of $\mu_S$. In the limit ${\mu}_S\to 0$ a leptonic U(1)
global symmetry is restored in the Lagrangian signifying exact conservation of 
lepton number that guarantees left-handed neutrinos to be massless \cite{naturalness,psb1,barr2}. In particular, a small and nonvanishing value of ${\mu}_S$
can be viewed as a slight breaking of the global $U(1)$ symmetry. Thus the smallness 
of ${\mu}_S$, desired in the TeV-scale inverse seesaw mechanism, which is 
 protected by the global symmetry in the 't Hooft sense
  \cite{gthooft}, is natural even though there is no dynamical understanding 
for such a small parameter. This view for the naturally small parameter ${\mu}_S$ being followed in the present work has been adopted in \cite{psb1} and by a number authors earlier pursuing inverse
 seesaw mechanism \cite{naturalness} although its interpretation through Higgs 
mechanism has been discussed in a model with extended gauge, fermion and Higgs
sectors \cite{ma2} and possibility of its radiative origin has been explored \cite{invrad}.

The physics underlying nonunitarity effects have been discussed at length
 in several recent papers \cite{antbnd,martinez,kanaya,kersten,malinsky2,altamelo,fdel,vander} where relevant
 formulas have been utilized. Although the PMNS matrix $U$ diagonalizes the
 light neutrino mass matrix of three generations where
 \ba
 U^{\dagger}m_{\nu}U^*&=&{\rm diag}(m_1,m_2,m_3)\equiv {\hat m}_{\nu},\label{udiag}
\ea
 the appropriate
 diagonalizing mixing matrix for the inverse seesaw matrix of eq.(\ref{inv}) is a 
$9\times9$ matrix $V$,
\ba
V^{\dagger}{\cal M}_{\nu}V^*&=&{\hat M}={\rm diag}(m_i,m_{R_j},m_{{\tilde
 R}_k}),\nonumber\\
&&(i,j, k=1,2,3),\label{vdiag}
 \ea
and this can be expressed in block partitions,
\ba
V&=&\begin{pmatrix}V_{3\times 3}&V_{3\times 6}\\V_{6\times 3}&V_{6\times 6}\end{pmatrix},\label{vblock}   
\ea
where the nonunitary $V_{3\times 3}$ matrix now represents the equivalent of
 the full PMNS matrix,
\ba
{\cal N}&\equiv& V_{3\times 3}\simeq (1-{1\over 2}XX^{\dagger})U\nonumber\\
&\simeq&(1-\eta)U.\label{matnu}
\ea 
Denoting the corresponding nine component eigenstate as $({\hat{\nu}}_i,N_i,
{\tilde N}_i)^T$, the six component heavy eigenstate as
$P^T=(N_1,N_2,N_3,{\tilde N}_1,{\tilde N}_2,{\tilde N}_3)^T$
and ${\cal K}\equiv V_{3\times 6}\simeq (0, X)V_{6\times 6}$, in  the leading
 order approximation in X, the light neutrino flavor eigenstate and the charged
 current Lagrangian in the mass basis are,
\ba
\nu^T &=& {\cal N}{\hat {\nu}}^T+{\cal K}P^T,\nonumber\\ 
{\cal L}_{CC}&=&-\frac{g_{2L}}{\sqrt 2}{\bar { l}}_L{\gamma}^{\mu}\nu
W^-_{\mu}+H.c.\nonumber\\
&\simeq&-\frac{g_{2L}}{\sqrt 2}{\bar { l}}_L\gamma^{\mu}({\cal N}{\hat
  {\nu}}^T+{\cal K}P^T)W^-_{\mu}+H.c.\label{ccnu}
\ea
The parameter $\eta=XX^{\dagger}/2$ characterizing nonunitarity of the
neutrino mixing matrix can have dramatic impact on leptonic CP-violation
and branching ratios for processes with lepton flavor violation (LFV),
\ba
{\cj}^{ij}_{\alpha\beta}&=&{\rm Im}({\cn}_{\alpha i} {\cn}_{\beta j} {\cn}^*_{\alpha
  j}{\cn}^*_{\beta i}),\nonumber\\
&\simeq&\cj+\Delta{\cj}^{ij}_{\alpha\beta},\label{cpnu}
\ea
where $\cj$ is the well known CP-violating parameter due to unitary PMNS matrix $U$ 
\ba
\cj &=& \cs{\cos^2\theta_{13}}\ca\sin\theta_{12}\sc\sa\sin\delta,\label{cpju}
\ea
and the nonunitarity
contributions are,
\ba  
\Delta{\cj}^{ij}_{\alpha\beta}&\simeq&-\sum_{\gamma =
  e,\mu,\tau}{\rm Im}(\eta_{\alpha\gamma}U_{\gamma i}U_{\beta j}U^*_{\alpha j}U^*_{\beta i}
\nonumber\\
&+&\eta_{\beta\gamma}U_{\alpha
  i}U_{\gamma j}U^*_{\alpha j}U^*_{\beta i}
\nonumber\\
&+&\eta_{\alpha\gamma}^*U_{\alpha i}U_{\beta j}U^*_{\gamma j}U^*_{\beta i}
\nonumber\\
&+&\eta_{\beta\gamma}^*U_{\alpha i}U_{\beta j}U^*_{\alpha j}U^*_{\gamma i}).
\label{cpj}
\ea 
Very recently $\sc$ has been measured \cite{T2K} to be small and nonvanishing although no
experimental information is available on the leptonic CP-phase $\delta$. Even
in the limiting case of vanishing unitarity CP-violation corresponding to $\sc \to
0,$ or $\delta \to 0, \pi$ for nonvanishing $\theta_{13}$
 nonunitarity effects caused due to $\eta$ may not vanish.    
In the modified charged current interaction in eq.(\ref{ccnu}), the heavy
neutrinos contribute to LFV decays with branching ratios \cite{ilakovac}
\ba
BR(l_{\alpha}\to l_{\beta }\gamma)&=&\frac{\alpha_w^3s_w^2m_{l_{\alpha}}^5}
{256\pi^2M_w^4\Gamma_{\alpha}}\nonumber\\
&&\times \left|\sum_{i=1}^6{\ck}_{\alpha i}{\ck}^*_{\beta
  i}I\left(\frac{m_{R_i}^2}{M_w^2}\right)\right|^2,\nonumber\\
I(x)&=& -\frac{2x^3+5x^2-x}{4(1-x)^3}-\frac{3x^3\ln x}{2(1-x)^4}.
\label{brlfv}
\ea
In eq.(\ref{brlfv}) the total decay width $\Gamma_{\alpha}$ for lepton species $l_{\alpha}$
with lifetime  $\tau_{\alpha}$ is evaluated using
$\Gamma_{\alpha} = \frac{{\hbar }}{ \tau_{\alpha}}$
where $\tau_{\mu}=(2.197019\pm 0.000021)\times 10^{-6}$~sec and
$\tau_{\tau}=(290.6\pm 1.0)\times 10^{-15}$~sec.

The matrix element
$({\ck}{\ck}^{\dagger})_{\alpha\beta}\propto {\eta}_{\alpha\beta}$ may lead to
significant lepton flavor violating (LFV) decays in the TeV scale
seesaw whereas LFV decays are drastically suppressed in Type-I seesaw in SO(10).
The procedure for estimating these effects has been outlined in
 ~\cite{psb1} which we follow. The  Dirac neutrino mass matrix at the TeV
scale which we derive in the next section is central to the determination of nonunitarity effects.

\section{V.RG EVOLUTION OF FERMION MASSES AND DETERMINATION OF $M_D$}

The determination of the Dirac neutrino mass matrix $M_D(M_{R^0})$ at the TeV seesaw
scale is done in three steps \cite{psb1}: (1) Derivation of RGEs for the
specific model and extrapolation of masses to the GUT-scale, (2) Fitting the
masses at the GUT scale and determination of $M_D(M_{GUT})$, (3) Determination of
 $M_D(M_{R^0})$ by top-down approach.\\

\begin{center}
{\bf A. RGEs and extrapolation to GUT scale}
\end{center}

At first RGEs
for Yukawa coupling matrices and fermion mass matrices are set up from which RGEs for mass
eigen values and CKM mixings are derived in the presence of $G_{2113}$
and $G_{2213}$ symmetries. RGEs in dynamical left-right breaking model has
been derived earlier \cite{akhme}.

Denoting $\Phi_{1,2}$ as the corresponding bidoublets under
 $G_{2213}$ they acquire VEVs
\ba
<\Phi_1>&=&\begin{pmatrix}v_u&0\\0&0\end{pmatrix},\nonumber\\
<\Phi_2>&=&\begin{pmatrix}0&0\\0&v_d\end{pmatrix}.\label{vevdef}  
\ea
Defining the mass matrices
\ba
M_u&=&Y_uv_u,~~M_D=Y_{\nu}v_u,~~M_d=Y_dv_d,\nonumber\\
M_e&=&Y_ev_d, M_R=y_{\chi} v_{\chi},\label{defmass} 
\ea 
we have derived the new RGEs in the presence of non-SUSY $G_{2113}$ and $G_{2213}$ gauge
symmetries for matrices $Y_i, M_i,i=u,d,e,N$, the mass eigenvalues
$m_i,i=u,c,t,d,s,b,e,\mu,\tau, N_1,N_2,N_3$, and  the CKM mixing matrix
elements as given in the Appendix B. 
We use the input values of running masses and quark mixings at 
 the electroweak scale as in refs.~\cite{nakamura,dp} 
and the resulting CKM matrix with the CKM Dirac phase  $\delta^q=1.20\pm 0.08$   
\ba
V_{\rm CKM}&=& \begin{pmatrix} 0.9742&0.2256&0.0013-0.0033i\\
-0.2255+0.0001i&0.9734&0.04155\\
0.0081-0.0032i&-0.0407-0.0007i&0.9991\end{pmatrix}.\label{vckz} 
\ea
We use RGEs of the standard model for $\mu=M_Z$ to $M_R^0=1$ TeV. With two Higgs 
doublets at $\mu \ge M_R^0$ we use the starting value of $\tan\beta=v_u/v_d=10$ 
at $\mu = 1$ TeV which evolves to reach the value  $\tan\beta\simeq 6.9$ at the 
GUT scale. Using the bottom-up approach discussed earlier \cite{dp} and the RGEs of Appendix B,  
the resulting quantities including the mass eigen values $m_i$ and the 
 $V_{CKM}$ at the GUT scale are \cite{conrad}, \\

\noindent{\large\bf{$\mu = {\rm M_{GUT}}$}:} 
\ba
m_e &=&0.48~{\rm MeV}, m_{\mu}={\rm 97.47}~{\rm MeV}, m_{\tau}=1.8814~
{\rm GeV},\nonumber\\
m_d&=&1.9~{\rm MeV},m_s=38.9~{\rm MeV}, m_b=1.4398~{\rm GeV},\nonumber\\ 
m_u&=&1.2~{\rm MeV}, m_c =0.264~{\rm GeV}, m_t = 83.04~{\rm GeV}, \label{eigenu}
\ea 
\vskip 0.2cm

\ba
V_{\rm CKM}(M_{GUT})&=& \begin{pmatrix} 0.9748& 0.2229& -0.0003-0.0034i\\
-0.2227-0.0001i& 0.9742& 0.0364\\
0.0084-0.0033i& -0.0354+0.0008i& 0.9993\end{pmatrix}.\label{vckmu} 
\ea 

\begin{center}
{\bf B. Determination of $M_D$}
\end{center}

With Higgs representations ${45}_H,{16}_H,{10}_H$, the ${\rm dim.}6$ operator \cite{psb1} 
\ba
{\bf \frac{f_{ij}}{M^2}16_i16_j{10}_H{45}_H{45}_H},\label{dim6}
\ea
with $M\simeq M_{Pl}$ or $M\simeq  M_{string}$ , is suppressed by $(M_U/M)^2
\simeq 10^{-3}-10^{-5}$ for GUT-scale VEV of ${45}_H$ and acts as an effective 
${126}_H$ operator to fit the fermion masses at the GUT scale where the
formulas for mass matrices are 
\ba
M_u &=& {G}_u + {F}, ~~M_d ={G}_d + {F},\nonumber\\
M_e &=& {G}_d -3 {F},~~M_D ={G}_u -3 {F}.
\label{fmU}
\ea
In eq.(\ref{fmU}) the matrices $ G_k=Y_k.16.16 <{10}^k_H>, k=u,d$ and $F$ are
derived from eq.(\ref{dim6}). 
Using a charged-lepton diagonal mass basis and eq.(\ref{eigenu}) and 
eq.(\ref{fmU}) we get,
\ba
M_e(M_{GUT})&=&{\rm diag}(0.0005,0.098,1.956)~{\rm GeV},\nonumber\\
 G_{d,ij}&=& 3F_{ij},~(i \neq j).\label{hdij}
\ea
Assuming for the sake of simplicity that the matrix $F$ is diagonal
 leads to the conclusion that the matrix $G_d$ is also diagonal. This gives 
relations between the diagonal elements which, in turn, determine the diagonal
matrices $F$ and $G_d$ completely\\
\ba
G_{d,ii} + F_{ii}&=&m_i,~(i=d,s,b),\nonumber\\
G_{d,jj} -3 F_{jj}&=&m_j,~(j=e,\mu,\tau),\label{hfrel}
\ea
\ba
F&=&{\rm diag}{1\over 4}(m_d-m_e, m_s-m_{\mu}, m_b-m_{\tau}),\nonumber\\
&=&{\rm diag}(3.75\times 10^{-4}, -0.0145, -0.3797)~{\rm GeV},\nonumber\\
G_d&=&{\rm diag}{1\over 4}(3m_d+m_e, 3m_s+m_{\mu}, 3m_b+m_{\tau}),\nonumber\\
&=&{\rm diag}(0.0016, 0.0544, 1.6709)~{\rm GeV},\label{FGd}
\ea
where we have used the RG extrapolated values of eq.(\ref{eigenu}).
Then using eq.(\ref{fmU}), eq.(\ref{FGd}) and the assumed basis gives the mass
matrices $M_u$ and $G_u$,
\ba
M_u(M_{GUT})&=&\begin{pmatrix}
0.0153&0.0615-0.0112i&0.1028-0.2706i\\
0.0615+0.0112i&0.3933&3.4270+0.0002i\\
0.1028+0.2706i&3.4270-0.0002&82.90\end{pmatrix}
~{\rm GeV},\label{MuU}
\ea
\ba
G_u(M_{GUT})&=&\begin{pmatrix}
0.0150&0.0615-0.0112i&0.1028-0.2706i \\
0.0615+0.0112i&0.4079&3.4270+0.0002i\\
0.1028+0.2706i&3.4270-0.0002i&83.01
\end{pmatrix}~{\rm GeV}.\label{GuU}
\ea
Now using eq.(\ref{FGd}) and eq.(\ref{GuU}) in eq.(\ref{fmU}) gives the Dirac
neutrino mass matrix $M_D$ at the GUT scale
\ba
M_D(M_{GUT})&=&\begin{pmatrix}
0.0139&0.0615-0.0112i&0.1029-0.2707i\\
0.0615+0.0112i&0.4519&3.4280+0.0002i\\
0.1029+0.2707i&3.4280-0.0002i&83.340
\end{pmatrix}~{\rm GeV}.
\ea 
We then use the RGE for $M_D$ given in Appendix A to evolve $M_D(M_{GUT})$ to
$M_D(M_{R^+})$ and then from $M_D(M_{R^+})$ to $M_D(M_{R^0})$ in two steps and
obtain,
\ba
M_D(M_{R^0})&=&\begin{pmatrix}
0.0151&0.0674-0.0113i&0.1030-0.2718i\\
0.0674+0.0113i&0.4758&3.4410+0.0002i\\
0.1030+0.2718i&3.4410-0.0002i&83.450
\end{pmatrix}~{\rm GeV}.\label{MD0}
\ea 
 
\section{VI.NONUNITARITY DEVIATIONS IN LEPTON MIXING MATRIX}

From eq.(\ref{matnu}) it is clear that any nonvanishing value of $\eta$ is a
measure of deviation from the unitarity of the PMNS matrix.
Using the TeV scale mass matrix for $M_D$ from eq.(\ref{MD0}) and assuming
\ba 
M_R&=&{\rm diag}(m_{R_1},m_{R_2},m_{R_3}),\label{MNdiag}
\ea
results in 
\ba
\eta &=& \frac{1}{2}X.X^{\dag}=M_DM_R^{-2}M_D^{\dag}/2,\nonumber\\
\eta_{\alpha\beta}&=&\frac{1}{2}\sum_{k=1,2,3}\frac{M_{D_{\alpha k}}M_{D_{\beta
    k}}^*}{m_{R_k}^2}.\label{etaele}
\ea

 For the sake of simplicity assuming degeneracy of RH neutrinos masses $m_R=m_{R_i}(i=1, 2, 3)$ gives,
\ba
\eta &=&\frac{1 {\rm GeV}^2}{m_R^2}
\begin{pmatrix}
0.0447&0.1937-0.4704i&4.4140-11.360i\\ 
0.1937+0.4704i&6.036&144.40-0.0002i\\ 
4.4140+11.360i&144.40+0.0002i&3488.0
\end{pmatrix}.\label{etaeq}
\ea

The deviations from unitarity in the leptonic mixing is constrained, for
 example, by deviations from universality tests in weak interactions, rare
 leptonic decays, invisible width of $Z$ boson and neutrino oscillation data.
The bounds derived at $90\%$ confidence level from the current data on the
 elements of the symmetric matrix are summarized in \cite{antbnd},
\ba
|\eta_{\tau\tau}|& \le & 2.7\times 10^{-3},~~|\eta_{\mu\mu}|\le 8.0\times 10^{-4},\nonumber\\
|\eta_{ee}|& \le &2.0\times 10^{-3},~~|\eta_{e\mu}|\le 3.5\times
 10^{-5},\nonumber\\
|\eta_{e\tau}| & \le & 8.0\times 10^{-3},~~|\eta_{\mu\tau}| \le  5.1\times
10^{-3}.\label{etabnd}
\ea
In the degenerate case the largest element in eq.(\ref{etaeq}) when compared with
$|\eta_{\tau\tau}|$ of eq.(\ref{etabnd}) gives the lower bound on the RH
neutrino mass,
\ba
m_R &\ge& 1.1366 ~{\rm TeV} \label{rhbnd},
\ea
which is only $7\%$ higher than the  SUSY $SO(10)$ bound $(m_R)_{SUSY}\ge
 1.06$ TeV \cite{psb1}.
Using this lower bound for other elements in eq.(\ref{etaeq}) yields
\ba
|\eta_{\mu\mu}|&\le& 4.672\times 10^{-6},\nonumber\\
|\eta_{ee}|& \le &3.460\times 10^{-8},~~|\eta_{e\mu}|\le 3.938\times
 10^{-7},\nonumber\\
|\eta_{e\tau}| & \le & 9.436\times 10^{-6},~~|\eta_{\mu\tau}| \le  1.1178\times
10^{-4}.\label{thbnd}
\ea
As in SUSY SO(10) \cite{psb1} , these predicted bounds are several orders lower than the current
experimental bounds and they might be reached provided corresponding LFV decays
are probed with much higher precision. 
But compared to SUSY $SO(10)$, in this model the upper bound is nearly $2$ times larger
for $|\eta_{\mu\tau}|$, $3$ times larger for $|\eta_{\mu\mu}|$, and nearly
 $40\%$ smaller in the case of  $|\eta_{e\tau}|$. It is interesting to note
 that in the the present non-SUSY $SO(10)$ model while some of the
 nonunitarity effects are comparable to the results of \cite{psb1}, others
 are distinctly different as shown in the next section.

\begin{figure}[htbp]
\centering
\includegraphics[width=0.30\textwidth, height=0.35\textheight, angle=-90]{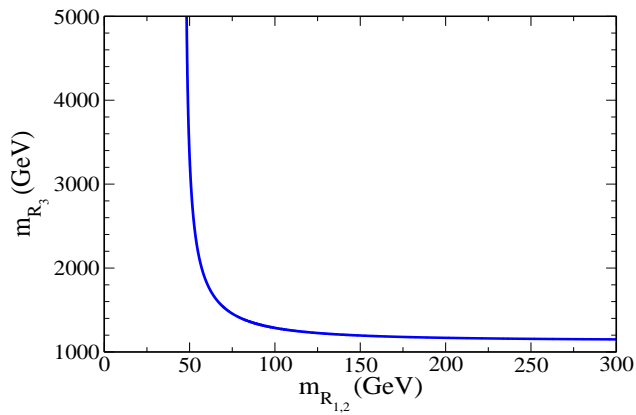}
\caption{Variation of the third generation RH neutrino mass $m_{R_3}$  as a function of
first or second generation neutrino mass $m_{R_1}$ or $m_{R_2}$ in the
partially degenerate case for which $m_{R_1}=m_{R_2}$. } 
\label{Fig3}
\end{figure}

We note in this model that when RH neutrino masses are nondegenerate, they are also constrained
by the experimental lower bound on $\eta_{\tau\tau}$ and
the corresponding relation obtained by saturating the bound is

\ba
\frac{1}{2}\left[\frac{0.0845}{m_{R_1}^2}+\frac{11.8405}{m_{R_2}^2}+\frac{6963.9}{m_{R_3}^2}\right]
 = 2.7\times 10^{-3},\label{costr}
\ea
where the numerators inside the square bracket are in GeV$^2$.  Using partial degeneracy, $m_{R_1}=m_{R_2}\neq m_{R_3}$ leads to the
relation between the RH neutrino masses as given in Table \ref{tabrel}. A plot
of $m_{R_3}$ vs. $m_{R_i}(i=1,2)$ is shown in Fig.\ref{Fig3} exhibiting increase
of  $m_{R_3}$ with decrease of $m_{R_i}$. The two asymptotes in the hyperbolic
curve are at $m_{R_1}=m_{R_2}\simeq 47$ GeV and $m_{R_3} \simeq 1136.6$ GeV. 

\begin{table*}
\begin{center}
\begin{tabular}{|c|c|c|c|}
\hline
$m_{R_{1,2}}$&$m_{R_3}$& $m_{R_{1,2}}$&$m_{R_3}$\\
(GeV)&(GeV)&(GeV)&(GeV)\\ \hline
48.0& 5572.83&500.0&1140.66\\ \hline
50.0& 3324.69& 600.0&1139.11\\ \hline
100.0& 1286.51&700.0&1138.18\\ \hline
150.0&1195.81&800.0&1137.57\\ \hline
200.0& 1168.32&900.0&1137.16\\ \hline
300.0& 1149.80&1000.0&1136.87\\ \hline
400.0&1143.53&1136.58&1136.58\\ \hline
\end{tabular}
\end{center}
\caption{Variation of third generation RH neutrino mass $m_{R_3}$ as a function of first or second
  generation RH neutrino mass in the partially degenerate case
  $m_{R_1}=m_{R_2}$ predicted by  nonunitarity through nonsupersymmetric  $SO(10)$.}
\label{tabrel}
\end{table*}

\begin{table*}
\begin{center}
\begin{tabular}{|c|c|c|c|c|c|c|c|c|}
\hline
$m_{R_1}=m_{R_2}$&$m_{R_3}$&$|\eta_{e\mu}|$&$\delta_{e\mu}$&$|\eta_{e\tau}|$&$
\delta_{e\tau}$&$|\eta_{\mu\tau}|$&$\delta_{\mu\tau}$\\
(GeV)&(GeV)&&&&&&\\ \hline
$1136$&$1136$&$3.938\times 10^{-7}$&$1.180$&$9.436\times 10^{-6}$&$1.20$&$1.118\times
10^{-4}$&$1.3\times 10^{-6}$ \\ \hline
$500$&$1141$&$4.222\times 10^{-7}$&$1.071$&$9.576\times 10^{-6}$&$1.166$&
$1.136\times 10^{-4}$&$2.0\times 10^{-4}$ \\ \hline
$100$&$1286$&$1.848\times 10^{-6}$&$0.308$&$1.687\times 10^{-5}$&$0.563$&
$1.691\times 10^{-4}$&$5.0\times 10^{-3}$ \\ \hline
$50$&$3325$&$6.733\times 10^{-6}$&$0.172$&$4.806\times 10^{-5}$&$0.202$&
$3.424\times 10^{-4}$&$1.0\times 10^{-2}$\\ \hline
\end{tabular}
\end{center}
\caption{Predictions of moduli and phases of nonunitarity parameters as a
  function of RH neutrino masses.}
\label{tabeta}
\end{table*}

\section{VII.ESTIMATIONS OF CP AND LEPTON FLAVOR VIOLATIONS AND DISCUSSIONS}

Two important physical applications of inverse seesaw are leptonic CP and flavor
violation effects reflected through the elements, both moduli and phases, of
the $\eta$-matrix and the relevant formulas have been discussed in Sec.4.  
The inverse seesaw formula of eq.(\ref{inv}) has three matrices out of which $M_D$
has been determined by fitting the charged fermion masses and mixings, but
since the~other two matrices, $M_N$ and ${mu}_S$,
can not be completely determined by using
the neutrino oscillation data alone, we make plausible assumptions.
In addition to the fully degenerate case we also examine consequences of
partial degeneracy with  $m_{R_1}=m_{R_2}$. 

From eq.(\ref{inv}), the nonunitary PMNS matrix $\cn = (1-\eta)U$ and the 
relation $m_{\nu}=\cn {\hat {m}}_{\nu}{\cn}^T$ give, 
\ba
\mu_S &=& X^{-1}\cn{\hat m}_{\nu}{\cn}^T(X^T)^{-1}.\label{mueq}
\ea 
We construct the unitary matrix $U$  using standard parameterization,

\ba
U&=&\begin{pmatrix}\uoo&\uot&\uoth\\ \uto&\utt&\utth\\ \utho&\utht&\uthth
\end{pmatrix},\label{umatr}
\ea  
and the neutrino oscillation
data at $3\sigma$ level \cite{T2K,fogli} and assuming hierarchical
neutrino masses,
\ba
\Delta m^2_{21}&=&(7.09 - 8.19)\times 10^{-5} {\rm eV}^2,\nonumber\\
\Delta m^2_{31}&=&(2.18 - 2.73)\times 10^{-3} {\rm eV}^2,\nonumber\\
\sin^2\theta_{12}&=& 0.27 - 0.36 ,\nonumber\\
\sin^2\theta_{23}&=& 0.39 - 0.64 ,\nonumber\\
\sin^2\theta_{13}&=&0.092\pm 0.06.\label{nudata}
\ea
We take the leptonic Dirac phase $\delta$
in the $U$ matrix to be zero for which the
predicted CP-violation from unitarity vanishes irrespective of the values of
$\theta_{13}$. We have also checked that inclusion of larger values of $\theta_{13} \simeq 8^o-9^o$ \cite{T2K} do not alter our results significantly. 
Similar results are obtained with $\delta = \pi$.

Taking the light neutrino mass eigen values
$m_1=0.001$ eV, $m_2=0.0088$ eV, $m_3=0.049$ eV, and the constructed $U$ matrix, 
we utilize  the $\eta$ matrix of eq.(\ref{etaeq}) for the degenerate case and 
eq.(\ref{matnu}) to obtain the nonunitary matrix $\cn$. Using eq.(\ref{mueq}) 
we also get the ${\mu}_S$ matrix. Once the matrices $\eta$ and $U$ are determined 
as discussed above and in Sec.4, the CP-violating parameters are computed using 
eq.(\ref{cpj}). Even though $U$ has no imaginary part because of assumed  vanishing value of its Dirac phase, CP-violation would arise from the 
imaginary parts of the corresponding components of $\eta$ matrix. We also estimate 
branching ratios for different LFV decay modes using eq.(\ref{brlfv}).

\begin{table*}
\begin{center}
\begin{tabular}{|c|c|c|c|c|c|c|c|c|c|}
\hline
$m_{R_1,R_2}$&$m_{R_3}$&$BR(\mu \to e\gamma)$&$BR(\tau
\to e\gamma)$&$BR(\tau \to \mu\gamma)$&$\Delta {\cj}^{12}_{e\mu}$&$\Delta
 {\cj}^{23}_{e\mu}$&$\Delta {\cj}^{23}_{\mu\tau}$&$\Delta
 {\cj}^{31}_{\mu\tau}$&$\Delta {\cj}^{12}_{\tau e}$\\
(GeV)&(GeV)&&&&&&&&\\ \hline
$1136$&$1136$&$2.0\times 10^{-16}$&$2.1\times 10^{-14}$&$3.0\times
 10^{-12}$&$-1.3\times 10^{-6}$&$-1.6\times 10^{-6}$&$1.6\times
10^{-6}$&$1.6\times 10^{-6}$&$4.0\times 10^{-6}$\\ \hline
$500$&$1140$&$2.0\times 10^{-16}$&$1.9\times 10^{-14}$&$2.7\times 10^{-12}$&
$-1.3\times 10^{-6}$&$-1.6\times 10^{-6}$&$1.6\times 10^{-6}$&$1.6\times 10^{-6}$&$4.0\times 10^{-6}$\\ \hline
$100$&$1286$&$1.4\times 10^{-15}$&$2.2\times 10^{-14}$&$2.2\times 10^{-12}
$&$-1.2\times 10^{-6}$&$-1.6\times 10^{-6}$&$2.2\times 10^{-6}$&$1.3\times 10^{-6}$&$4.1\times 10^{-6}$ \\ \hline
$50$&$3325$&$1.1\times 10^{-14}$&$1.1\times 10^{-13}$&$5.5\times 10^{-12}$&$
-1.0\times 10^{-6}$&$-1.8\times 10^{-6}$&$4.1\times 10^{-6}$&$7.4\times 10^{-7}$&$4.3\times 10^{-6}$ \\ \hline
\end{tabular}
\end{center}
\caption{Nonunitarity predictions of leptonic CP-violating parameters and
  branching ratios for lepton flavor violating decays $\mu \to e\gamma$, 
  $\tau \to e\gamma$, and $\tau \to \mu\gamma$ as a function RH neutrino masses}
\label{tabcpv}
\end{table*}
\begin{table*}
\begin{center}
\begin{tabular}{|l|c|c|}
\hline
$m_{R_{1,2}}$&$m_{R_3}$& Mass eigenvalues $\mu_{S_i}$\\
(GeV)&(GeV)&(MeV)\\ \hline
50& 3324.7& ($2.4583$, $3.23\times 10^{-3}$, $1.18\times 10^{-6}$) \\ \hline
100& 1286.5&($8.0423$, $2.60\times 10^{-3}$, $1.07\times 10^{-6}$) \\ \hline
500& 1140.7& ($199.37$, $5.29\times 10^{-2}$, $1.05\times 10^{-6}$) \\ \hline
1136.6& 1136.6&($1030.0$, $2.72\times 10^{-1}$, $1.04\times 10^{-6}$) \\ \hline
\end{tabular}
\end{center}
\caption{Mass eigen values of ${\bf \mu}_S$ signifying masses of singlet fermions
  predicted by the inverse seesaw in $SO(10)$} 
\label{eigenmu}
\end{table*}

For the degenerate case with $m_R=1.1366$ TeV  we get

\ba
{\bf\mu}_S&=&\begin{pmatrix}
0.9932-0.0124i&-0.1908+0.0022i&0.0066-0.0033i\\
-0.1908+0.0022i&0.0370-0.0004i&-0.0013+0.0006i\\
0.0066-0.0033i&-0.0013+0.0006i&0.00003-0.00004i\\
\end{pmatrix}{\rm GeV}, \nonumber
\ea
\ba
\Delta {\cj}^{12}_{e\mu}&=&-1.3082\times 10^{-6},\nonumber\\
\Delta {\cj}^{23}_{e\mu}&=&-1.5573\times 10^{-6},\nonumber\\
\Delta {\cj}^{23}_{\mu\tau}&=&1.5574\times 10^{-6},\nonumber\\
\Delta {\cj}^{31}_{\mu\tau}&=&1.5572\times 10^{-6},\nonumber\\
\Delta {\cj}^{12}_{\tau e}&=&4.0144\times 10^{-6},\label{cpvsol1}
\ea
and the branching ratios
\ba
BR(\mu \to e\gamma)&=&2.0025\times 10^{-16},\nonumber\\
BR(\tau \to e\gamma)&=&2.1586\times 10^{-14},\nonumber\\
BR(\tau \to \mu\gamma)&=&3.0290\times 10^{-12}.\label{lfvbr}
\ea

Thus we find that in this non-SUSY $SO(10)$ model  for the degenerate RH neutrino masses, like the SUSY SO(10) prediction \cite{psb1}, 
although all the five CP violating parameters are just one order
smaller than the corresponding parameter in the quark sector where ${\cj
}_{CKM}=(3.05^{+0.19}_{-0.20})\times 10^{-5}$, there are certain
quantitative differences. The magnitudes of predicted CP-violations for all the five
parameters  in the non-SUSY $SO(10)$ model are reduced by
nearly $50\%$ compared to their corresponding SUSY $SO(10)$ values.

When compared with the predicted values in SUSY $SO(10)$ \cite{psb1} the present results on
branching ratios satisfy,
\ba
\frac{{BR(\mu\to e\gamma)}_{\rm susy}}{{BR(\mu\to e\gamma)}_{\rm
    non-susy}}&\simeq& {3\over 2} ,\nonumber\\
\frac{{BR(\tau\to e\gamma)}_{\rm susy}}{{BR(\tau\to e\gamma)}_{\rm
    non-susy}}&\simeq& {5} ,\nonumber\\
\frac{{BR(\tau\to \mu\gamma)}_{\rm susy}}{{BR(\tau\to \mu\gamma)}_{\rm
    non-susy}}&\simeq& {2\over 3} .\label{compbr}
\ea
which can be tested by next generation experiments on LFV decays.

Our predictions for the partially degenerate RH neutrinos on different
elements ${\eta}_{\alpha\beta}$ and their phases are given in Table
\ref{tabeta} and those for CP-violating parameters $\Delta {\cj}_{\alpha\beta}^{ij}$
and branching ratios are summarized in Table \ref{tabcpv}.

Compared to the predictions in the degenerate case,  $|\eta_{\mu\tau}|\simeq 10^{-4}$,
$\delta_{\mu\tau}\simeq 10^{-6}$, for the partially degenerate case we find that 
while $|\eta_{\mu\tau}|$ is of the same order, but $\delta_{\mu\tau}\simeq 10^{-2}, 
10^{-3}$ and $10^{-4}$  for $m_{R_{1,2}}=50$ GeV, $100$ GeV, and $500$ GeV,
respectively. These parameters enter into the neutrino oscillation
probability in the ``golden channel'' \cite{martinez},
\ba 
P_{\mu\tau}&\simeq&4|\eta_{\mu\tau}|^2+4s_{23}^2c_{23}^2
\sin^2\left(\frac{\Delta  m_{31}^2L}{4E}\right)\nonumber\\
&-&4|\eta_{\mu\tau}|\sin\delta_{\mu\tau}\sin 2\theta_{23}\sin\left(\frac{\Delta
  m_{31}^2L}{4E}\right).\label{golden}
\ea
leading to the CP-asymmetry,
\ba
{\cal A}^{CP}_{\mu\tau}&=& \frac{P_{\mu\tau}-P_{{\bar {\mu}}{{\bar
	{\tau}}}}}{P_{\mu\tau}+P_{{\bar {\mu}}{{\bar {\tau}}}}}\nonumber\\ 
&\simeq& \frac{-4|\eta_{\mu\tau}|\sin\delta_{\mu\tau}}
{\sin 2\theta_{23}\sin\left(\frac{\Delta
  m_{31}^2L}{4E}\right)}.\label{asydef}
\ea
when the first term in  eq.(\ref{golden}) is much smaller compared to the 
other two terms.  Our results in the partial degenerate case satisfiy  the
condition that gives eq.(\ref{asydef}) from eq.(\ref{golden}). 
The nonunitarity CP violating effects are predicted
to be much more pronounced by noting that the strength of the third term in
eq.(\ref{golden}) is enhanced    
by $100-10,000$ times compared to the prediction in the degenerate case. 
Crucial to this prediction is our constraint eq.(\ref{costr}) between RH
neutrino masses which plays an important role in estimating 
the phase of $\eta_{\mu\tau}$ in the partially degenerate case that takes into
account the increasing behavior of $m_{R_3}$ for decreasing values of $m_{R_1}=m_{R_2}$.

Among other significant differences in the model predictions are 
Br($\mu\to e\gamma$) values higher by two orders or by one order for
$m_{R_1}=m_{R_2}=50$ GeV or $100-500$ GeV while  
Br($\tau\to e\gamma$) is predicted to be one order lower for 
the  RH neutrino masses $m_{R_1}=m_{R_2}=100-1180$~GeV.  
 Presently the experimental limits on branching ratios are Br($\mu\to e\gamma$)$\le
2.4\times 10^{-12}$ \cite{MEG},
Br($\tau\to e\gamma$)$\le 1.2\times 10^{-7}$ \cite{Belle}, and  Br($\tau\to \mu\gamma$)$\le
4.5\times 10^{-8}$ \cite{Belle}. The projected reach of future sensitivities are
up to Br($\tau\to e\gamma$) $\sim 10^{-9}$, Br($\tau\to \mu\gamma$)$\sim
10^{-9}$, but Br($\mu\to e\gamma$)$\sim 10^{-14}$\cite{joaquim,PRISM}.

In Table \ref{eigenmu} we show predictions of mass eigen values of the
${\bf\mu}_S$ matrix that signifies masses of three fermion singlets
$S_i\,(i=1,2,3)$ for degenerate and partially degenerate cases of RH neutrino
masses. These mass eigen values are noted to vary starting from the lightest
$\sim 1$ eV to the heaviest $\sim 1$ GeV which may have interesting
phenomenological consequences that needs further investigation. It is to be
noted that the smallest mass eigen value is also predicted directly by the
inverse seesaw formula from the TeV scale value of $(M_D)_{33}\sim 100$ GeV
in a manner similar to the Type-I seesaw case.
 
We have also examined the consequences of quasidegenerate light neutrino masses 
expected to manifest through tritium beta decay or neutrinoless double beta decay 
searches. For example with $m_1=0.09923$ eV, $m_2=0.09965$ eV, and  $m_3=0.1110$ eV, 
which are consistent with neutrino oscillation data, the three eigen values of the 
resulting ${\mu}_{S}$ matrix are  ${\mu}^{(i)}_{S} = (30.110\,{\rm GeV}, 1.2\,{\rm MeV}, 
20.6\,{\rm eV})$ with three pairs of heavy pseudo Dirac neutrinos having almost 
degenerate  masses (1151.7, 1121.6) GeV, (1139.5, 1139.5) GeV, and (1136.5, 1136.5) GeV. 
The predictions for LFV decays, CP-violating parameters and the nonunitarity effects 
are similar to the case of the degenerate pseudo Dirac neutrinos with hierarchical light 
neutrino masses as discussed above. However, the heaviest eigen value of the fermion singlet mass matrix increases to $\mu^{(1)}_S \simeq 30$ GeV compared to the corresponding value of  $\mu^{(1)}_S \simeq 1$ GeV 
in the hierarchical case of light neutrinos as shown in Table \ref{eigenmu}.

The introduction of three additional fermion singlets under $SO(10)$ needed for the implementation of inverse seesaw mechanism may be argued to be a limitation of the related GUT models. For that matter, the other $SO(10)$ models of refs. \cite{smirnov,pr,psb1,barr1,barr2,psb2} have utilized these singlets to obtain different interesting results. More recently, the superpartners of two out of these three fermion singlets have been demonstrated to be acting as components of inelastic dark matter \cite{psb3}. 
There is another $SO(10)-$ based radiative inverse seesaw  model               
which has been designed to explain the smallness of the ${\mu}_S$ parameter with the symmetry breaking chain $SO(10)\to SU(5)\times U(1)_{\chi} \to SM\times U(1)_{\chi}$ where more nonstandard fermions and singlets have been found to be necessary \cite{invrad}. These indicate the popularity of $SO(10)$-singlet fermion models inspite of the stated limitation.   

In this respect the $E_6$  \cite{gursey} or $SU(3)^3$ \cite{trini,cpwsu33,masu33} type GUT models do not have this limitation as they contain the necessary fermion singlets within their fundamental representations but they also  contain a number of additional nonstandard fermions. In the absence of any experimental data on the masses of these additional fermions at low energies, the determination of the Dirac neutrino mass matrix from fermion mass fits at the GUT scale using the bottom-up approach adopted here is not possible.  
As one major objective of the present work is the prediction on the lifetimes of gauge boson mediated proton decay $p \to e^+\pi^0$ on which dedicated search experiments are ongoing \cite{nishino}, $SU(3)^3$ type of GUTs do not serve this objective as the corresponding decays are 
suppressed \cite{trini,cpwsu33}. This model may be important if, ultimately, 
 proton decay search experiments  observe a very large lower limit on the 
lifetime.

It has been also argued that because of large size of Higgs representations such as ${210}_H$ and ${126}_H$
      needed in  $SO(10)$  models employing type-I and type-II seesaw mechanisms, GUT-threshold corrections may give rise to larger uncertainties in
 $\sin^2\theta_{W}$ predictions and associated mass scale(s)~
\cite{sher} . Counter examples of this result in SO(10) having Pati-Salam intermediate symmetry ($SU(2)_L\times SU(2)_R\times SU(4)_C\times D\equiv G_{224D}$) 
with unbroken D-Parity have been derived with exactly vanishing GUT-threshold corrections on $\sin^2\theta_W$ as well as on the intermediate  scale \cite{parpat}. It has been also shown how threshold corrections can be reduced substantially in other $SO(10)$ models with naturally plausible constraint that all superheavy components of a $SO(10)$ Higgs representation are degenerate in 
masses \cite{mp,lmpr}.                
Noting that the Higgs representation ${126}_H$
is needed for the implementation of the type-I and type-II seesaw mechanisms, and the inverse seesaw needs comparatively much smaller Higgs representation like ${16}_H$, the possibilities of threshold uncertainties are expected to be correspondingly reduced in our models. In particular our minimal Model-I contains neither of the larger Higgs representations 
 ${210}_H$ and ${126}_H$; it requires                
only the smaller representations $ {45}_H, {16}_H$ and~${10}_{H_1},{10}_{H_2}$.
Also in case of the Model-II and its extension, the  GUT-threshold effects due to superheavy components of Higgs representations ${210}_H, {16}_H$ and ${10}_{H_1},{10}_{H_2}$~are expected to be substantially reduced  compared to the 
$SO(10)$ model of ref.\cite{lmpr} with $G_{2213D}$ intermediate symmetry because of the absence of the large representation ${126}_H$. The maximal value of proton lifetime is found to increase by a factor 2(4) due to GUT threshold effects in  our Model-I (Model-II) over the two-loop predictions. 

Regarding other possibilities of inverse seesaw motivated non-SUSY $SO(10)$, we find that the minimal single-step breaking scenario to the TeV scale gauge symmetry , $SO(10)\to G_{2113}$, is ruled out by renormalization group and coupling unification constraints. One of the two-step breaking chains, 
$SO(10)\to G_{224D} \to G_{2113}$ gives a low value of the unification scale $M_U = 10^{14.7}$ GeV whereas $SO(10)\to G_{214} \to G_{2113}$ also yields almost 
 similar value, $M_U = 10^{14.8}$ GeV where we have used $SU(2)_L\times U(1)_R\times SU(4)_C \equiv G_{214}$. The third remaining chain, $SO(10)\to G_{224} \to G_{2113}$, where D-parity is broken at the GUT scale, gives $M_U=10^{15.15}$ GeV. Thus all the three minimal chains at two-loop level are ruled out by the existing lower bouind on proton lifetime \cite{nishino}. As the large representation ${126}_H$ is absent in these models, the GUT-threshold effects \cite{lmpr} are smaller in the corresponding minimal models than the required values to make them compatible with the lower limit on proton lifetime unless the splitting among the superheavy components is too large. In view of these, the minimal Model-I turns out to be the best among all possible single and two-step breaking minimal models of $SO(10)$ with the TeV scale $G_{2113}$ gauge symmetry. 

One of the appealing features which have been noted \cite{dpar1} in  $SO(10)$ breaking chains under the category of Model-I is that they do not have the cosmological domain wall problem \cite{shafi} because of spontaneous breaking of  
D-Parity along with the gauge symmetry at the GUT scale. When this criteria is included while searching for equally good models, there are only two possible chains with three step breakings and only one chain with four step breaking to the TeV-scale symmetry $G_{2113}$. However, if utilization of large Higgs representations is excluded, the minimal Model-I emerges to be unique from among all
possible $SO(10)$ breaking chains.   
Investigation of prospects for these longer symmetry breaking chains along with others which is beyond the scope of the present work will be addressed elsewhere.    

\section{VIII.SUMMARY AND CONCLUSION}

We have investigated the prospects of inducting TeV-scale inverse seesaw
mechanism for neutrino masses into nonsupersymmetric $SO(10)$ grand unification
and found that it can be successfully implemented with a low-mass $Z^{\prime}$
gauge boson
accessible to experimental detection at LHC and planned accelerators. By
setting up RGEs in the presence of $G_{2213}$  and $G_{2113}$ gauge
symmetries we have extrapolated fermion masses and mixings to the GUT scale
using bottom-up approach and determined the Dirac neutrino mass matrix from a successful fit at the GUT
scale. We have found a relation between the RH neutrino masses 
which, in the partially degenerate case, predicts the third generation RH
neutrino mass to increase substantially with
the decrease of first or second generation RH neutrino masses. Although the 
predicted branching ratios in the case of degenerate RH neutrinos show
less than one order variations from the corresponding SUSY SO(10) predictions,
in the partially degenerate case, the branching ratio Br$(\mu\to e\gamma)$ is predicted to be 
larger by  $1-2$ orders while Br $(\tau\to e\gamma)$ is  predicted
to be lower by one order for all values of allowed RH neutrino masses.
For the nonunitarity matrix element ${\eta}_{\mu\tau}$ an important model prediction is its enhanced phase
$\delta_{\mu\tau}$ larger by $2-4$ orders which is expected to play a dominant role in the
experimental detection of the nonunitary CP-violation effects at neutrino
factories. We have also shown that the models accommodate quasidegenerate light
neutrino masses relevant for neutrinoless double beta decay or the tritium beta 
decay searches with predictions on the LFV, CP-violation, and nonunitarity effects 
similar to the case of hierarchical light neutrinos and degenerate pseudo Dirac neutrinos while the heaviest mass of the fermion singlets increases from 
 $\mu^{(1)}_S \simeq 1$ GeV to $\mu^{(1)}_S \simeq 30$ GeV. 

Interestingly, the two-loop prediction on proton lifetime in the  minimal model (Model-I) turns out to be  $[\tau_p(p\to
e^+\pi^0)]_{\rm max.}=7\times 10^{34}$ Yrs. which increases by a factor of $2$  when GUT threshold effects are included. While providing a possibility of verification of the underlying GUT hypothesis, this offers another opportunity
 for testing the minimal model by ongoing search experiments regarding its validity
or falsifiability. We have also identified this model to be the best among all  involving single or two-step breakings of $SO(10)$ to the TeV scale gauge symmetry $G_{2113}$ which is essential for low mass $Z^{\prime}$ and prominent nonunitarity effects. But if large Higgs representations are excluded from symmetry breakings, the minimal Model-I turns out to be a unique model among all pssible $SO(10)$ symmetry breaking chains.

Too fast proton decay in another model ( Model-II) has been shown to be evaded by a simple extension where some of the predictions on $\tau_p$  should be within the reach of future experiments. On the other hand, if the actual proton lifetime  is too large, this is also shown to be accommodated in model extensions along with associated nonunitarity and lepton flavor violation effects with the prospect of detection of a color octet scalar at accelerator energies.

In conclusion we find that induction of TeV-scale inverse seesaw mechanism into nonsupersymmetric $SO(10)$ predicts pronounced nonunitarity  
and CP-violating effects  measurable at accelerator energies and neutrino factories for hierarchical as well as partially degenerate spectra of light neutrino masses. In the TeV scale inverse seesaw mechanism motivated GUT model, these effects are mainly due to predominance of the Dirac 
neutrino mass matrix in $SO(10)$ because of its underlying quark-lepton
symmetry and this holds even if only an experimentally verifiable low-mass $Z^{\prime}$ gauge boson is present as one of the smoking gun signatures of asymptotic parity restoration.

\vskip 0.5cm
\par\noindent{\bf {ACKNOWLEDGMENTS}}\\

\noindent{M.K.P. thanks  Harish-Chandra Research Institute for a visiting position. The authors thank Sandhya Choubey for discussion.}\\

\section{APPENDIX A}
In the standard notation of two-loop evolution equations for gauge couplings, 
\ba
\frac{\mu{\partial g}_i}{{\partial
    \mu}}&=\frac{1}{16\pi^2}a_ig_i^3+\frac{1}{(16\pi^2)^2}\sum_jb_{ij} g_i^3 g_j^2,
  \nonumber\\
\ea
the one- and two-loop beta function coefficients are given in Table \ref{tabgi}.
 We have noted a small contribution of $U(1)_R\times U(1)_{B-L}$
mixing effect \cite{mix} especially in the case of Model-I. 

\begin{table*}
\begin{center}
\begin{tabular}{c|c|c|c|}
\hline
 $MODEL$&$Symmetry$&${a_i}$&${b_{ij}}$\\
         &&&(GeV)\\ \hline
$I,II$&$G_{213}$&$\begin{pmatrix}-19/6,41/10,-7\end{pmatrix}$&$\begin{pmatrix}
199/50,27/10,44/5\\9/10,35/6,12\\11/10,9/2,-26\end{pmatrix}$ \\ \hline
$I,II$&$G_{2113}$&$\begin{pmatrix}-3,53/12,33/8,-7\end{pmatrix}$&$\begin{pmatrix}8,1,3/2,12\\3,17/4,15/8,12\\9/2,15/8,65/16,4\\9/2,3/2,1/2,-26\end{pmatrix}$ \\ \hline
$I$&$G_{2213}$&$\begin{pmatrix}-8/3,-13/6,17/4,-7\end{pmatrix}$&$\begin{pmatrix}37/3,6,3/2,12\\6,143/6,9/4,12\\9/2,27/4,37/8,4\\9/2,9/2,1/2,-26\end{pmatrix}$ \\ \hline
$II$&$G_{2213D}$&$\begin{pmatrix}-13/6,-13/6,17/4,-7\end{pmatrix}$&$\begin{pmatrix}143/6,6,9/4,12\\6,143/6,9/4,12\\27/4,27/4,23/4,4\\9/2,9/2,1/2,-26\end{pmatrix}$\\ \hline
\end{tabular}
\end{center}
\caption{One-loop and two-loop beta function coefficients for gauge coupling
  evolutions in Model-I and Model-II described in the text taking the second
  Higgs doublet mass at $1$ TeV} 
\label{tabgi}
\end{table*}
\section{APPENDIX B}
Each of the two $SO(10)$ models we have considered for inverse seesaw has two types of
nonstandard gauge symmetries, $G_{2213}$ or $G_{2213D}$ and $G_{2113}$.
Here we derive RGEs for running Yukawa and fermion mass matrices from which
,following the earlier approach \cite{dp}, we derive RGEs for the mass
eigenvalues and mixing angles.
We define the rescaled $\beta$- functions 
\ba
 16\pi^2\mu\frac{{\partial {F}_i}}{{\partial {\mu}}}=
\beta_{F_i}.\label{B1} 
\ea
With $G_{2113}$ symmetry the scalar field $\Phi_d(2,1/2,0,1)$ through its VEV $v_d$ gives 
masses to down quarks and charged leptons while $\Phi_u(2,-1/2,0,1)$ through its VEV $v_u$
gives Dirac masses to up quarks and neutrinos. These fields are embedded into
separate bi-doublets in the presence of $G_{2213}$ and their vacuum structure
has been specified in Sec.4. We have derived the beta functions for
RG evolution of Yukawa matrices ($Y_i$), fermion mass matrices ($M_i$), and 
the vacuum expectation values ($v_{u,d}$). The rescaled 
beta functions  are given below in both cases.\\

\noindent{\large \bf {$G_{2113}$ Symmetry:}}
\ba
\beta_{Y_u}&=&[{3\over 2}Y_uY_u^{\dag}+ {1\over 2}Y_dY_d^{\dag}+T_u-\sum_i
  C_i^qg_i^2]Y_u,
\nonumber\\
\beta_{Y_d}&=&[{3\over 2}Y_dY_d^{\dag}+{1\over 2}Y_uY_u^{\dag}+T_d-
\sum_i C_i^qg_i^2]Y_d,  
\nonumber\\
\beta_{Y_{\nu}}&=&[{3\over 2}Y_{\nu}Y_{\nu}^{\dag}+{1\over 2} Y_eY_e^{\dag}+T_u-\sum_i
  C_i^{l}g_i^2]Y_{\nu}, \nonumber\\
\beta_{Y_e}&=&[{3\over 2}Y_eY_e^{\dag}+{1\over 2} Y_{\nu}Y_{\nu}^{\dag}+T_d-\sum_i
  C_i^{l}g_i^2]Y_e, \nonumber\\
\beta_{M_u} &=&
[{3\over 2}Y_uY_u^{\dag}+{1\over 2}Y_dY_d^{\dag}-\sum_i{\tilde C}^q_ig_i^2]M_u,
\nonumber\\
\beta_{M_d} &=&
[{3\over 2}Y_dY_d^{\dag}+ {1\over 2}  Y_uY_u^{\dag}-\sum_i{\tilde C}^q_ig_i^2]M_d, \nonumber\\
\beta_{M_D}&=&[{3\over 2}Y_{\nu}uY_{\nu}^{\dag}+{1\over 2}Y_eY_e^{\dag}-\sum_i{\tilde C}^l_ig_i^2]M_D,\nonumber\\
\beta_{M_e}&=&[{3\over 2}  Y_euY_e^{\dag}+{1\over 2}  Y_{\nu}Y_{\nu}^{\dag}-\sum_i{\tilde
    C}^l_ig_i^2]M_e,\nonumber\\
\ea
where the beta-functions for VEVs are
\ba
\beta_{v_u}&=&[\sum_iC^{v}_ig_i^2 -T_u]v_u,\nonumber\\
\beta_{v_d}&=&[\sum_iC^{v}_ig_i^2 -T_d]v_d,\label{matrge}
\ea
with
\ba 
T_u={\rm Tr}(3Y_u^{\dag}Y_u+Y_{\nu}^{\dag}Y_{\nu}),~ T_d={\rm Tr}(3Y_d^{\dag}Y_d+
Y_e^{\dag}Y_e).\label{tutd}
\ea
The parameters occurring in these equations, and also in eq.(\ref{eigenbeta}) and eq.(
\ref{ckmgen}) given below are
\ba
a &=&{3\over 2},~b={1\over 2},~a^{\prime}=b^{\prime}=0,\nonumber\\
C^q_i&=&(9/4,3/4,1/4,8),~C^l_i=(9/4,3/4,9/4,0),\nonumber\\
{\tilde C}^q_i&=&(0,0,1/4,8),{\tilde C}^l_i = (0,0,9/4,0),~C^{v}_i=(9/4,3/4,0,0),~(i=2L,1R,BL,3C).\label{paramat1}
\ea

\noindent{\large \bf {$G_{2213}$ Symmetry:}}\\

Following definitions of Sec.4 in the presence of left-right symmetry. the
rescaled beta functions for RGEs of the Yukawa and fermion mass matrices are
\ba
\beta_{Y_u}&=&(Y_uY_u^{\dag}+Y_dY_d^{\dag})Y_u+Y_u(Y_u^{\dag}Y_u+Y_d^{\dag}Y_d)
+{T}_uY_u+{\hat
  T}_1Y_d-
\sum_i   C_i^qg_i^2Y_u, \nonumber\\
\beta_{Y_d}&=&(Y_dY_d^{\dag}+ Y_uY_u^{\dag})Y_d+Y_d(Y_d^{\dag}Y_d+
Y_u^{\dag}Y_u)+{T}_dY_d+{\hat T}_2Y_u -\sum_i  C_i^qg_i^2Y_d, 
\nonumber\\
\beta_{Y_{\nu}}&=& (Y_{\nu}Y_{\nu}^{\dag}+Y_eY_e^{\dag})Y_{\nu}+
 Y_{\nu}(Y_{\nu}^{\dag}Y_{\nu}+Y_e^{\dag}Y_e)+{T}_uY_{\nu}+{\hat T}_1Y_e
-\sum_i C_i^{l}g_i^2Y_{\nu}, \nonumber\\
\beta_{Y_e}&=&(Y_eY_e^{\dag}+Y_{\nu}Y_{\nu}^{\dag})Y_e+
Y_e(Y_e^{\dag}Y_e+Y_{\nu}^{\dag}Y_{\nu})+{T}_dY_e+
{\hat T}_2Y_{\nu }-\sum_i  C_i^lg_i^2Y_e, \nonumber
\ea
\ba
\beta_{M_u} &=&
(Y_uY_u^{\dag}+Y_dY_d^{\dag})M_u+M_u(Y_u^{\dag}Y_u+Y_d^{\dag}Y_d)
-\sum_i{\tilde C}^q_ig_i^2M_u+{\hat T}_1\tan\beta M_d,\nonumber\\
\beta_{M_d} &=&
(Y_dY_d^{\dag}+Y_uY_u^{\dag})M_d+M_d(Y_d^{\dag}Y_d+Y_u^{\dag}Y_u)     
-\sum_i{\tilde C}^q_ig_i^2]M_d+{{\hat T}_2\over
  \tan\beta}M_u,\nonumber\\
\beta_{M_D}&=&(Y_{\nu}uY_{\nu}^{\dag}+ Y_eY_e^{\dag})M_D+M_D
(Y_{\nu}^{\dag}Y_{\nu}+Y_e^{\dag}Y_e)
-\sum_i{\tilde C}^l_ig_i^2M_D+{\hat T}_1\tan\beta M_e,\nonumber\\
\beta_{M_e}&=&(Y_eY_e^{\dag}+Y_{\nu}Y_{\nu}^{\dag})M_e+
M_e(Y_e^{\dag}Y_e+Y_{\nu}^{\dag}Y_{\nu})
-\sum_i{\tilde C}^l_ig_i^2M_e
+{{\hat T}_2\over\tan\beta}M_D,\label{g2213}
\ea
where the rescaled beta functions for VEVs $\beta_{v_u}, \beta_{v_d}$ are the
same as in eq.(\ref{matrge}) with different
coefficients $C_i^v$  defined below and 
functions $T_u$ and $T_d$ are the same as in 
eq.(\ref{tutd}). Other two traces entering in this case are
\ba
{\hat T}_1&=& {\rm Tr}(3Y_d^{\dag}Y_u+Y_e^{\dag}Y_{\nu}),\nonumber\\
{\hat T}_2&=&{\rm Tr}(3Y_u^{\dag}Y_d+Y_{\nu}^{\dag}Y_e).\label{hat12}
\ea
The parameters occurring in these equations and also in eq.(\ref{eigenbeta}) and
eq.(\ref{ckmgen}) given below are 
\ba
a&=&b=2,~a^{\prime}=b^{\prime}=1,\nonumber\\
C^q_i&=&(9/4,9/4,1/4,8),~C^l_i=(9/4,9/4,9/4,0),~{\tilde C}^q_i=(0,0,1/4,8),\nonumber\\
{\tilde C}^l_i&=&(0,0,9/4,0),~C_i^{v}=(9/4,9/4,0,0),~(i=2L,2R,BL,3C).
\label{paramat2} 
\ea
Then following the procedure described in \cite{dp}, and using the definition of
parameters in the two different mass ranges, given above we obtain RGEs for mass
eigenvalues and elements of CKM mixing matrix $V_{\alpha\beta}$ which can be expressed in the 
generalized form for both cases,\\

\noindent{\large \bf{Mass Eigenvalues:}}\\
\ba
\beta_{m_i}&=&\left[-\sum_k{\tilde C}_k^{(q)}g_k^2+ a y_i^2+
   2 b \sum_{j=d,s,b}|V_{uj}|^2y_j^2+a^{\prime}{{{\hat T}_1\tan\beta}\over m_i}\sum_{j=d,s,b}|V_{uj}|^2m_j\right]m_i,~i=u,c,t~,\nonumber\\
\beta_{m_i}&=&\left[-\sum_k{\tilde C}_k^{(q)}g_k^2+ ay_i^2+
   2b\sum_{j=u,c,t}|V_{dj}|^2y_j^2+b^{\prime}{{\hat T}_2\over {\tan\beta m_i}}\sum_{j=u,c,t}|V_{dj}|^2m_j\right]m_i,~i=d,s,b~,\nonumber\\
\beta_{m_i}&=&\left[-\sum_k {\tilde C}_k^{(l)}g_k^2+ay_i^2+
    2b\sum_{j=N_1,N_2,N_3}y_j^2+b^{\prime}{{\hat T}_2\over {\tan\beta m_i}}\sum_{j=N_1,N_2,N_3}m_j\right]m_i,~i=e,\mu,\tau~,\nonumber\\
\beta_{m_i}&=&\left[-\sum_k{\tilde C}_k^{(l)}g_k^2+ ay_i^2
  +a^{\prime}{{{\hat T}_1\tan\beta}\over
    m_i}\sum_{j=e,\mu,\tau}m_j\right]m_i,~i=N_1,N_2,N_3~.\label{eigenbeta}
\ea
\noindent{\large\bf{CKM Matrix Elements:}}\\
\ba
\beta_{V_{\alpha\beta}} &=&\sum_{\gamma=u,c,t; \gamma\neq \alpha}\left[
 a^{\prime}\frac{{\hat T}_1\tan\beta}{m_{\alpha}-m_{\gamma}}(V{\hat M}_dV^{\dag})_{\alpha
\gamma}+\frac{b}{v_d^2}\frac{m_{\alpha}^2+m_{\gamma}^2}{m_{\alpha}^2-
m_{\gamma}^2}(V{\hat M}_d^2V^{\dag})_{\alpha\gamma}\right]V_{\gamma\beta}\nonumber\\
&&-\sum_{\gamma=d,s,b; \gamma\neq \beta}V_{\alpha\gamma}\left[b^{\prime}\frac{
{\hat T}_2}
{\tan\beta(m_{\gamma}-m_{\beta})}(V^{\dag}{\hat M}_uV)_{\gamma\beta}+\frac{b}{v_u^2}\frac{m_{\gamma}^2+m_{\beta}^2}{m_{\gamma}^2-
m_{\beta}^2}(V^{\dag}{\hat M}_u^2V)_{\gamma\beta}\right].\label{ckmgen}
\ea
Then using third generation dominance, the beta functions for all the 9
elements are easily obtained for respective mass ranges where in addition to
the parameters in the respective cases in eq.(\ref{paramat1}) and
eq.(\ref{paramat2}), $a^{\prime}=b^{\prime}=0$ in the mass range $M_{R^0}\to
M_{R^+}$  with $G_{2113}$ symmetry, but   $a^{\prime}=b^{\prime}=1$ in the mass range     $M_{R^+}\to
M_U$ with $G_{2213}$ or $G_{2213D}$ symmetry and, in the latter case, the
nonvanishing traces ${\hat T}_{1,2}$ are easily
evaluated in the mass basis.

\end{document}